\def\be{\begin{equation}}
\def\ee{\end{equation}}
\def\bea{\begin{eqnarray}}
\def\eea{\end{eqnarray}}
\def\ba{\begin{array}{l l}}
\def\ea{\end{array}}

\documentclass[aip,jap,reprint,graphicx]{revtex4-1} 
\usepackage{amssymb,amsmath}
\usepackage[usenames,dvips]{color}
\usepackage{graphicx}
\usepackage{xcolor}
\usepackage{dcolumn}
\usepackage{bm}

\begin{document}



\title{Topological signatures of the coexistence of antiferromagnetism and odd-parity spin-triplet superconductivity} 

\author{Maria Teresa Mercaldo}
\affiliation{Dipartimento di Fisica ``E. R. Caianiello", Universit\`a di Salerno, IT-84084 Fisciano (SA), Italy}
\author{Panagiotis Kotetes}
\affiliation{Niels Bohr Institute, University of Copenhagen, 2100 Copenhagen, Denmark}
\author{Mario Cuoco}
\affiliation{CNR-SPIN, IT-84084 Fisciano (SA), Italy}
\affiliation{Dipartimento di Fisica ``E. R. Caianiello", Universit\`a di Salerno, IT-84084 Fisciano (SA), Italy}

\begin{abstract}
Strongly correlated systems exhibit a rich phenomenology due to the antagonism of a diversity of ordered phases. The aftermath of this interplay can lead to a coexistence which takes place at a microscopic level, or, a phase separation in which non-overlapping single-order domains extend throughout the material. In most cases it appears experimentally challenging to disentangle the two scenarios, unless, there exist robust and measurable properties particular to only one of the two types of coexistence. This is for instance the case when the type of coexistence decides on the appearance of topologically protected excitations, such as, Majorana fermions. In this work, we explore a concrete example falling into this category of systems, and specifically, we investigate one-dimensional odd-parity spin-triplet superconductors in the presence of antiferromagnetism. We determine the symmetry conditions for the occurrence of Majorana edge states and explore their response to variations of the strength and orientation of the antiferromagnetic field $\bm{M}$, as well as, the spin structure of the Cooper pairs controlled by the so-called $\bm{d}$-vector. 
\end{abstract}

\maketitle

\section{Introduction}

A common aspect of interacting quantum systems is the close competition, or even the coexistence, of dif\-fe\-rent types of ordered phases. Spin-triplet superconductivity (STSC) is a paradigmatic phase of matter in this respect, as it often appears in proximity to ferromagnetic quantum phase transitions~\cite{Fay,Pfleiderer}. For example, this situation is encountered in heavy fermions superconductors, i.e., $\mathrm{UGe}_2$, URhGe and $\mathrm{UIr}_2$, as well as, materials on the verge of compe\-ting magnetic instabilities, e.g., ruthe\-na\-tes~\cite{Maeno,Maeno2}. This is a quite remarkable result as, in this manner, STSC allows Cooper pairing to take place even in the pre\-sen\-ce of strong intrinsic or extrinsic ferromagnetic fields. This is in contrast to spin-singlet supercon\-ducti\-vi\-ty (SC), which is generally expected to survive only when the magnetization profile yields a zero net moment as, for instance, in the case of antiferromagnetic (AFM) orde\-ring. Then, the interplay of magnetism and SC encodes useful information regarding the Cooper-pairing type and glue. 

In spite of the significance of the above interplay, key questions, such as, whether magnetism and SC overlap in the same spatial domain, co\-exist, compete, or result in new phases, remain fairly unanswered for a large class of quantum materials, e.g. cuprates, iron-based and heavy fermion superconductors, the Bechgaard salts and others. In fact, for the latter family of superconductors, the Cooper pairs are very likely to have a spin-triplet symmetry, as suggested by the so far observed upper critical fields and exi\-sting NMR measurements~\cite{Lee97}. Additional experimental evidences~\cite{Kornilov2004,Lee2005} also indicate a region of coexistence between spin-triplet SC and antiferromagnetism. The possible realization of such a coexistence phase has motivated various theoretical stu\-dies relating to an emergent SO(4) symmetry~\cite{Podolsky2004}, the interface energy at the boundary of superconducting and AFM domains~\cite{Zhang99}, the multi-orbital STSC with AFM order \cite{Spalek}, as well as inhomogeneous STSC and AFM phase~\cite{Zhang2006}.

Apart from the great attention that spin-triplet pai\-ring enjoys for its foundational role in unconventional SC~\cite{sigrist91,Maeno}, it also constitutes a pole of attraction for applications in the field of superconducting spintronics~\cite{Linder2015} and topological quantum computation based on Majorana fermions~\cite{Yakovenko,SRFL08,RSFL10,qi11,tanaka12,Beenakker13,Flensberg2012,SatoAndo2017}. The interest in super\-con\-ducting spin\-tronics typically emphasizes the role of ferromagnetism. Unconventional phenomena along this direction are often encountered, such as the emergent spin-orbital interaction between the superconducting order parameter (OP) and the interface ferromagnetism~\cite{Gentile13,Terrade16}, the breakdown of the bulk-boundary correspondence~\cite{Mercaldo16,Mercaldo18}, and the anomalous magnetic~\cite{Romano13,Romano16} and spin-charge current~\cite{Romano17} effects in the proximity between chiral or helical $p$-wave and spin-singlet superconductors. Moreover, within the framework of quantum systems that combine topological and conventional forms of order, the very recent observation of an anomalous coexistence phase of AFM and SC in monolayer FeTe grown on a topological insulator~\cite{Manna2017}, reinforces the idea that unexpected quantum effects may take place in the presence of Dirac physics, electron pai\-ring and magnetism~\cite{Brzezicki2018}.    

In this paper we address the occurrence of zero-energy Majorana edge states in one-dimensional spin-triplet superconductors with coexisting AFM order. We demonstrate how the strength and orientation of the AFM field $\bm{M}$ and the spin structure of the Cooper pairs, as given by the $\bm{d}$-vector, affect the topological phase diagram. In particular, a chiral-symmetry-protected pair of Majorana fermions per edge are remarkably robust if the magnetization $\bm{M}$ has an easy plane configuration and the $\bm{d}$-vector is perpendicular to it. On the other hand, we find that a topological phase harbouring a single Majorana fermion per edge turns out to be more fragile and significantly dependent on the variation of the orientation of $\bm{M}$. Therefore, our analysis allows to identify distinct topological behaviors when $\bm{M}$ and $\bm{d}$ are oriented either in a parallel or a perpendicular fashion.

\begin{figure}
\includegraphics[width=0.48\columnwidth]{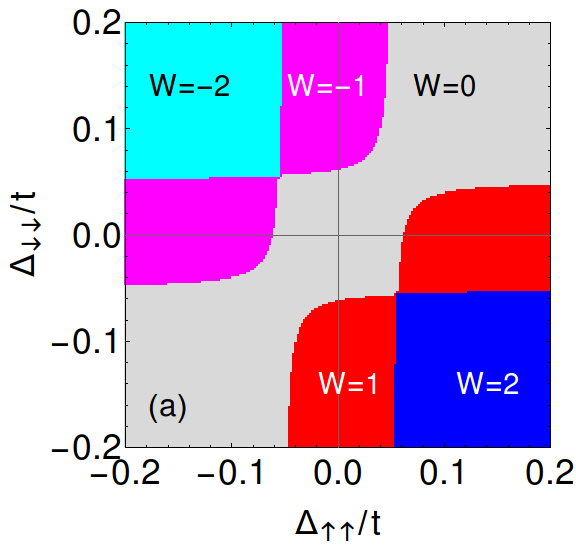}
\includegraphics[width=0.48\columnwidth]{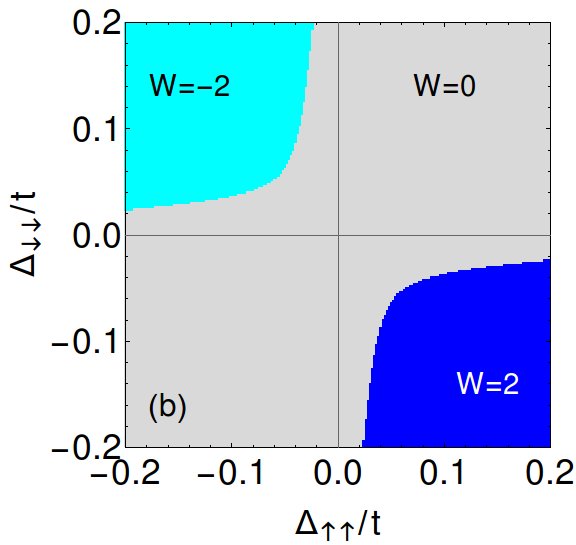}
\caption{Topological phase diagram in the $[\Delta_{\uparrow\uparrow},\Delta_{\downarrow\downarrow}]$ plane for representative values of the AFM-field strength $|\bm{M}|$ and the chemical potential $\mu$ (namely $|\bm{M}|/t$=0.15 and $\mu/t=0.1$), and two different orientations of the AFM magnetization in the $yz$ plane, i.e. $\bm{M}$ almost parallel to $z$ axis ($\theta=0.1\pi$) in (a) and to $y$ axis in (b). We find topological phases with either 1 or 2 Majorana fermions per edge, which are mainly accessible when $\Delta_{\uparrow\uparrow}$ and $\Delta_{\downarrow\downarrow}$ have opposite sign, i.e. the $\bm{d}$-vector primarily points along the $x$ direction. The sign of the winding number also depends on the sign of the difference between $\Delta_{\uparrow\uparrow}$ and $\Delta_{\downarrow\downarrow}$. In addition, with $\Delta_{\uparrow\uparrow}$ and $\Delta_{\downarrow\downarrow}$ having the same sign, only topological phases with a single Majorana fermion per edge can be obtained.}
\label{fig:fig1}
\end{figure}

\section{Model and Methodology}

We model the one-dimensional $p$-wave superconductor in the presence of an AFM order with a Bogoliubov-de Gennes (BdG) lattice Hamiltonian in $k$ space as, ${\cal H}=\frac{1}{2}\sum_k\Psi_k^{\dag}\widehat{{\cal H}}_k\Psi_k$, where
\bea
\widehat{{\cal H}}_k &=&-\mu\tau_z-2t\cos(ka)\tau_z\rho_xe^{-ika\rho_z}-\rho_z\bm{M}\cdot\bm{\sigma}\nonumber \\
&+&\big(\tau_+ \bm{d}_k+\tau_- \bm{d}^*_k\big)\cdot\rho_xe^{-ika\rho_z}\bm{\sigma}\,,
\label{fin}
\eea


\noindent with $\bm{\sigma}$, $\bm{\rho}$, and $\bm{\tau}$ being the Pauli matrices for spin, magnetic sublattice ($A$ and $B$) and particle-hole channels, respectively. The raising and lowering operators in the latter space read $\tau_{\pm}=(\tau_x\pm i\tau_y)/2$. Above we have employed the multi-component spinor
\bea
\Psi_k^{\dag}&=&\big(\psi_{kA\uparrow}^{\dag},\,\psi_{kA\downarrow}^{\dag},\,\psi_{kB\uparrow}^{\dag},\,\psi_{kB\downarrow}^{\dag},\,\nonumber\\
&&\,\psi_{-kA\downarrow},\,-\psi_{-kA\uparrow},\,\psi_{-kB\downarrow},\,-\psi_{-kB\uparrow}\big)
\eea

\noindent In addition, we introduced the form factor $\bm{d}_k=2\sin(ka)\bm{d}$, where $a$ defines the lattice constant and $k$ takes values in the reduced Brillouin zone $(-\frac{\pi}{2a},\frac{\pi}{2a}]$, since the length of the magnetic unit cell is equal to $2a$. We also introduced the AFM OP $\bm{M}$, which has opposite orientation and equal amplitude on the two sublattices. The $\bm{d}$-vector is generally complex and sets the spin orien\-ta\-tion of the OP for STSC. It is convenient to introduce the matrix OP in spin-space $\{\uparrow,\downarrow\}$, $\widehat{\Delta}=\bm{d}\cdot\bm{\sigma}\sigma_y$: $\Delta_{\uparrow\uparrow,\downarrow\downarrow}=d_y\pm id_x$ and $\Delta_{\uparrow\downarrow}=-id_z$. The $\bm{d}$-vector components are then related to the spin-triplet configurations having zero spin projection along the corresponding symmetry axis.

For the present analysis, we assume that the superconducting OP has an easy $xy$ spin-plane for the $\bm{d}$-vector and allow for a generic orientation along a given direction in the plane, as indicated by the angle $\alpha$, i.e. ${\bm{d}}=d(i\cos\alpha,\sin\alpha,0)$. Here, we consider a $\bm{d}$-vector a\-ri\-sing from a real amplitude of $\Delta_{\uparrow\uparrow,\downarrow\downarrow}$. Such a spin structure for the pai\-ring can be achieved by introducing an effective separable four-fermion inte\-rac\-tion in the $p$-wave channel with potentials $V_x\neq V_y$ in the $xy$ spin-plane and $V_z=0$ along the $z$ spin-axis. For the given $\bm{d}$-vector configuration, there exist orientations of the AFM OP $\bm{M}$, for which the BdG Hamiltonian can reside in the BDI symmetry class~\cite{Altland,KitaevClassi,Ryu} exhi\-bi\-ting chiral, time-reversal and charge-conjugation symmetries. As a matter of fact, for any given choice which restricts the orientation of the $\bm{d}$-vector to the $xy$ spin-plane, the Hamiltonian posseses a chiral symmetry effected by the ope\-ra\-tor $\widehat{\Pi}=\tau_x \sigma_x$, as long as $\bm{M}$ lies in the $yz$ plane. It is convenient to parametrize the AFM exchange field as $\bm{M}=(0,M\sin\theta,M\cos\theta)$, with the angle $\theta$ setting the orientation and $M=|\bm{M}|$ the modulus of the respective OP in the $yz$ spin-plane.

Since the chiral-symmetry operator anticommutes with $\widehat{{\cal H}}_k$, the Hamiltonian can be cast into an off-diagonal form with antidiagonal blocks expressed in terms of a matrix $\hat{A}_{k}$ and its hermitian conjugate $\hat{A}_{k}^{\dag}$. This is achieved by employing a unitary transformation $\widehat{U}$ rotating the basis in the eigenbasis of $\widehat{\Pi}$. 
Hence, the determinant $\det{\hat{A}_{k}}$, can be put in a complex polar form $z_k= |\det{\hat{A}_{k}}| \exp[i \varphi_k]$ and, as long as the eigenvalues of $\hat{A}_{k}$ are non-zero, it can be used to obtain the winding number $W$ by evaluating its trajectory in the complex plane. The winding number is defined as 
\bea
W=\frac{1}{2\pi i}\int_0^{2\pi}d\varphi
\eea

\noindent and constitutes the topological invariant of the system. Therefore, it predicts the number of zero-energy Majorana fermions which appear at each edge of the corresponding open system. This bulk-boundary correspondence follows from an index theorem, cf. Ref.~\onlinecite{Sato2011PRB}, which can be employed to express the winding number as $W=n_{+}-n_{-}$, with $n_{+}$ and $n_{-}$ being the number of eigenstates associated to the eigenvalues $\pm1$ of the chiral operator $\widehat{\Pi}$. 

Below we analyze the topological behavior of the one-dimensional $p$-wave superconductor for various AFM configurations by determining the win\-ding number in the parameters space and also by solving the real space BdG equations on an open chain with finite length. Note that in this work we do not solve the BdG equation in a self-consistent manner, thus, neither accounting for the backreaction of the AFM order onto the $p$-wave OP nor for any possible edge-reorganization effects. Such a self-consistent treatment, additionally emphasizing the possibility of the bulk-boundary correspondence breakdown, has been previously carried out in Ref.~\onlinecite{Mercaldo16} for the case of a STSC in the presence of a ferromagnetic field. We expect that similar effects may be also become important here when the superconducting coherence length becomes comparable to the AFM unit cell size.

\section{Results}

We start by considering some limiting cases with the $\bm{d}$-vector oriented along the $y$ and $x$ axis. When dealing with a pure $d_y$ spin-state (i.e. $\Delta_{\uparrow\uparrow}=\Delta_{\downarrow\downarrow}=\Delta$), then, the superconductor is always topologically trivial, except for an AFM magnetization which is along $z$, thus, perpendicular to $d_y$. This result can be immediately obtained by inspection of the dependence of the winding number $W$ on $\Delta_{\uparrow\uparrow}$ and $\Delta_{\downarrow\downarrow}$. It turns out that, for $\theta\neq0,\pi$, the imaginary part of $W$ is proportional to the difference $\Delta_{\uparrow\uparrow}-\Delta_{\downarrow\downarrow}\propto d_x$ so that there is no possibility to achieve any winding in the $d_y$ configuration when the OPs have equal amplitude and sign (see also Fig.~\ref{fig:fig1}). For the specific case of having the AFM magnetization along $z$ and a $d_y$ spin configuration, the $z$ component of the electron spin is a good quantum number and the problem can be separated into two independent blocks containing only $\Delta_{\uparrow\uparrow}$ or $\Delta_{\downarrow\downarrow}$. The structure of the winding number, in this circumstance, resembles that of the spinless $p$-wave Kitaev chain but with a residual dependence on the amplitude of the OP, that, however, returns a non-trivial topological phase in a large portion of the $(\mu,t)$ phase space.

A completely different topological behavior is obtained for a $p$-wave superconductor with a $d_x$ spin configuration. In this limit $\Delta_{\uparrow\uparrow}=-\Delta_{\downarrow\downarrow}=\Delta$ and, as expected, the topological phases do not depend on the orientation of the AFM magnetization in the $yz$ plane. Moreover, we find that for this configuration, the winding number can acquire only the values $\{0,\pm2\}$ (note the diagonals of Fig.~\ref{fig:fig1}). The phase with two Majorana fermions per edge is in this case quite robust with respect to changes in the amplitude of the exchange field and the electron filling controlled by $\mu$. In addition, our analytics show that in the regime where $M<|\mu|$, the occurrence of 2 Majorana fermions per edge can be observed up to va\-ni\-shing amplitudes of $\Delta$. This is because in this case the SC OP does not enter the topological criterion, which entails that a non-trivial phase is reached when $4t^2+M^2>\mu^2$ is sa\-tis\-fied. Instead, if $M>|\mu|$ the winding number yields $|W|=2$ only when $4\Delta^2+\mu^2>M^2$.

\begin{figure}
\includegraphics[width=0.8\columnwidth]{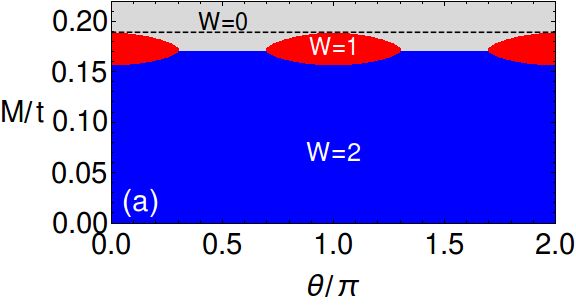}
\includegraphics[width=0.8\columnwidth]{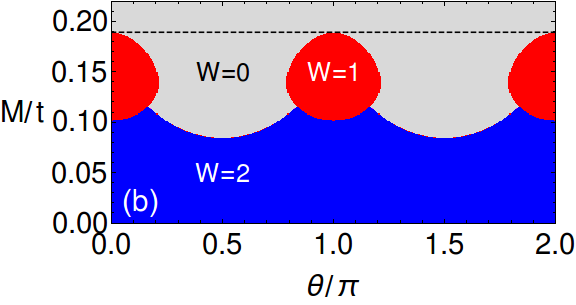}
\caption{Evolution of the topological phases in terms of the amplitude ($M$) and orientation ($\theta$) of the AFM field $\bm{M}$ for two different configurations of the SC OP $\bm{d}$-vector. Specifically, $\mu=0.1t$, $\Delta_{\uparrow\uparrow}=0.08t$ and in (a) $\Delta_{\downarrow\downarrow}=-0.06t$ while in (b) $\Delta_{\downarrow\downarrow}=-0.01t$. The dashed line indicates the separation between the region with 2 and 0 Majorana fermions per edge for the case of $\Delta_{\uparrow\uparrow}=-\Delta_{\downarrow\downarrow}$. We observe that the phases with a single-Majorana per edge never occur in the $d_x$ spin configuration, and that the extension and occurrence of the area with $W=1$ is interrelated to the amplitude mismatch between $\Delta_{\uparrow\uparrow}$ and $\Delta_{\downarrow\downarrow}$.}
\label{fig:fig2}
\end{figure}

\begin{figure}
\includegraphics[width=0.9\columnwidth]{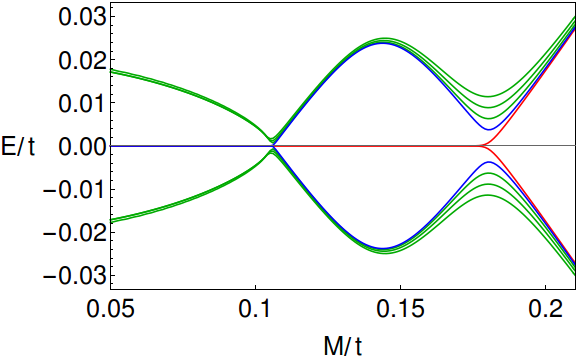}
\caption{Evolution of the low energy states close to the Fermi level for an open chain with $N=2000$ sites as a function of the exchange amplitude $M$ at a given orientation $\theta=0.1 \pi$ of the AFM magnetization corresponding to a vertical trajectory in the panel (b) of Fig.~\ref{fig:fig2}. We observe that a series of gap closings and reopenings is obtained when crossing the boundary between the regions with 2,1, and 0 Majorana edge states.}
\label{fig:fig3}
\end{figure}

When moving away from the $d_x$ and $d_y$ symmetric configurations the aspect of the topological phase diagram is significantly modified, leading to a new phase with a single Majorana fermion per edge (Fig.~\ref{fig:fig1}), which exhibits a marked dependence on the orientation of the AFM spin moments. Indeed, as shown in Fig.~\ref{fig:fig2}, the $|W|=1$ region arises when moving away from the limit $\Delta_{\uparrow\uparrow}=-\Delta_{\downarrow\downarrow}$ but only nearby specific orientations of the AFM magnetization. In addition, its stability region generally extends in size, away from the high symmetry points at $\theta=0,\pi$, by increasing the amplitude mismatch between $\Delta_{\uparrow\uparrow}$ and $\Delta_{\downarrow\downarrow}$. The evolution of the electronic spectra for an open chain for a given trajectory in the phase space (Fig. \ref{fig:fig3}) confirms the occurrence of 2- and 1-Majorana fermions per edge with a gap closing and re-opening associated with the topological transitions across the phase boundary.

\section{Conclusions}

In conclusion we have shown that chiral-symmetry-protected Majorana edge states can occur when $p$-wave superconductivity coexists with antiferromagnetism. Depending on the relative orientation of the $\bm{d}$-vector and the AFM field $\bm{M}$, edge states with one or two Majorana fermions per edge are expected to be observed. 
Although the topological protection holds only for an easy plane orientation of the AFM magnetization for a given $\bm{d}$-vector configuration, such an anisotropy is commonly encountered in antiferromagnets and the vanishing of the net magnetization avoids to have misalignments due to stray fields as in ferromagnets. Another important remark is that the topological phases are expected to be robust against spatial inhomogeneities of the superconducting OPs as far as they  are not leading to gap clo\-sings. Finally, since it is challenging to expe\-ri\-men\-tally identify the microscopic coexistence phase, we point out that, due to the distinctive properties of the Majorana fermions, the detection of topologically protected edge states can be employed to assess the degree of overlap of the two orders and the character of the magnetic and superconducting spin structure in the coexisting phase. 

%



\begin{thebibliography}{34}%
\makeatletter
\providecommand \@ifxundefined [1]{%
 \@ifx{#1\undefined}
}%
\providecommand \@ifnum [1]{%
 \ifnum #1\expandafter \@firstoftwo
 \else \expandafter \@secondoftwo
 \fi
}%
\providecommand \@ifx [1]{%
 \ifx #1\expandafter \@firstoftwo
 \else \expandafter \@secondoftwo
 \fi
}%
\providecommand \natexlab [1]{#1}%
\providecommand \enquote  [1]{``#1''}%
\providecommand \bibnamefont  [1]{#1}%
\providecommand \bibfnamefont [1]{#1}%
\providecommand \citenamefont [1]{#1}%
\providecommand \href@noop [0]{\@secondoftwo}%
\providecommand \href [0]{\begingroup \@sanitize@url \@href}%
\providecommand \@href[1]{\@@startlink{#1}\@@href}%
\providecommand \@@href[1]{\endgroup#1\@@endlink}%
\providecommand \@sanitize@url [0]{\catcode `\\12\catcode `\$12\catcode
  `\&12\catcode `\#12\catcode `\^12\catcode `\_12\catcode `\%12\relax}%
\providecommand \@@startlink[1]{}%
\providecommand \@@endlink[0]{}%
\providecommand \url  [0]{\begingroup\@sanitize@url \@url }%
\providecommand \@url [1]{\endgroup\@href {#1}{\urlprefix }}%
\providecommand \urlprefix  [0]{URL }%
\providecommand \Eprint [0]{\href }%
\providecommand \doibase [0]{http://dx.doi.org/}%
\providecommand \selectlanguage [0]{\@gobble}%
\providecommand \bibinfo  [0]{\@secondoftwo}%
\providecommand \bibfield  [0]{\@secondoftwo}%
\providecommand \translation [1]{[#1]}%
\providecommand \BibitemOpen [0]{}%
\providecommand \bibitemStop [0]{}%
\providecommand \bibitemNoStop [0]{.\EOS\space}%
\providecommand \EOS [0]{\spacefactor3000\relax}%
\providecommand \BibitemShut  [1]{\csname bibitem#1\endcsname}%
\let\auto@bib@innerbib\@empty
\bibitem [{\citenamefont {Fay}\ and\ \citenamefont {Appel}(1980)}]{Fay}%
  \BibitemOpen
  \bibfield  {author} {\bibinfo {author} {\bibfnamefont {D.}~\bibnamefont
  {Fay}}\ and\ \bibinfo {author} {\bibfnamefont {J.}~\bibnamefont {Appel}},\
  }\href@noop {} {\bibfield  {journal} {\bibinfo  {journal} {Phys. Rev. B}\
  }\textbf {\bibinfo {volume} {22}},\ \bibinfo {pages} {3173} (\bibinfo {year}
  {1980})}\BibitemShut {NoStop}%
\bibitem [{\citenamefont {Pfleiderer}(2009)}]{Pfleiderer}%
  \BibitemOpen
  \bibfield  {author} {\bibinfo {author} {\bibfnamefont {C.}~\bibnamefont
  {Pfleiderer}},\ }\href@noop {} {\bibfield  {journal} {\bibinfo  {journal}
  {Rev. Mod. Phys.}\ }\textbf {\bibinfo {volume} {81}},\ \bibinfo {pages}
  {1551} (\bibinfo {year} {2009})}\BibitemShut {NoStop}%
\bibitem [{\citenamefont {Maeno}\ \emph {et~al.}(1994)\citenamefont {Maeno},
  \citenamefont {Hashimoto}, \citenamefont {Yoshida}, \citenamefont
  {Nishizaki}, \citenamefont {Fujita}, \citenamefont {Bednorz},\ and\
  \citenamefont {Lichtenberg}}]{Maeno}%
  \BibitemOpen
  \bibfield  {author} {\bibinfo {author} {\bibfnamefont {Y.}~\bibnamefont
  {Maeno}}, \bibinfo {author} {\bibfnamefont {H.}~\bibnamefont {Hashimoto}},
  \bibinfo {author} {\bibfnamefont {K.}~\bibnamefont {Yoshida}}, \bibinfo
  {author} {\bibfnamefont {S.}~\bibnamefont {Nishizaki}}, \bibinfo {author}
  {\bibfnamefont {T.}~\bibnamefont {Fujita}}, \bibinfo {author} {\bibfnamefont
  {J.~G.}\ \bibnamefont {Bednorz}}, \ and\ \bibinfo {author} {\bibfnamefont
  {F.}~\bibnamefont {Lichtenberg}},\ }\href@noop {} {\bibfield  {journal}
  {\bibinfo  {journal} {Nature}\ }\textbf {\bibinfo {volume} {372}},\ \bibinfo
  {pages} {532} (\bibinfo {year} {1994})}\BibitemShut {NoStop}%
\bibitem [{\citenamefont {Mackenzie}\ and\ \citenamefont
  {Maeno}(2003)}]{Maeno2}%
  \BibitemOpen
  \bibfield  {author} {\bibinfo {author} {\bibfnamefont {A.~P.}\ \bibnamefont
  {Mackenzie}}\ and\ \bibinfo {author} {\bibfnamefont {Y.}~\bibnamefont
  {Maeno}},\ }\href@noop {} {\bibfield  {journal} {\bibinfo  {journal} {Rev.
  Mod. Phys.}\ }\textbf {\bibinfo {volume} {75}},\ \bibinfo {pages} {657}
  (\bibinfo {year} {2003})}\BibitemShut {NoStop}%
\bibitem [{\citenamefont {Lee}\ \emph {et~al.}(1997)\citenamefont {Lee},
  \citenamefont {Naughton}, \citenamefont {Danner},\ and\ \citenamefont
  {Chaikin}}]{Lee97}%
  \BibitemOpen
  \bibfield  {author} {\bibinfo {author} {\bibfnamefont {I.~J.}\ \bibnamefont
  {Lee}}, \bibinfo {author} {\bibfnamefont {M.~J.}\ \bibnamefont {Naughton}},
  \bibinfo {author} {\bibfnamefont {G.}~\bibnamefont {Danner}}, \ and\ \bibinfo
  {author} {\bibfnamefont {P.~M.}\ \bibnamefont {Chaikin}},\ }\href@noop {}
  {\bibfield  {journal} {\bibinfo  {journal} {Phys. Rev. Lett.}\ }\textbf
  {\bibinfo {volume} {78}},\ \bibinfo {pages} {3555} (\bibinfo {year}
  {1997})}\BibitemShut {NoStop}%
\bibitem [{\citenamefont {Kornilov}\ \emph {et~al.}(2004)\citenamefont
  {Kornilov}, \citenamefont {Pudalov}, \citenamefont {Kitaoka}, \citenamefont
  {Ishida}, \citenamefont {q.~Zheng}, \citenamefont {Mito},\ and\ \citenamefont
  {Qualls}}]{Kornilov2004}%
  \BibitemOpen
  \bibfield  {author} {\bibinfo {author} {\bibfnamefont {A.~V.}\ \bibnamefont
  {Kornilov}}, \bibinfo {author} {\bibfnamefont {V.~M.}\ \bibnamefont
  {Pudalov}}, \bibinfo {author} {\bibfnamefont {Y.}~\bibnamefont {Kitaoka}},
  \bibinfo {author} {\bibfnamefont {K.}~\bibnamefont {Ishida}}, \bibinfo
  {author} {\bibfnamefont {G.}~\bibnamefont {q.~Zheng}}, \bibinfo {author}
  {\bibfnamefont {T.}~\bibnamefont {Mito}}, \ and\ \bibinfo {author}
  {\bibfnamefont {J.~S.}\ \bibnamefont {Qualls}},\ }\href@noop {} {\bibfield
  {journal} {\bibinfo  {journal} {Phys. Rev. B}\ }\textbf {\bibinfo {volume}
  {81}},\ \bibinfo {pages} {224404} (\bibinfo {year} {2004})}\BibitemShut
  {NoStop}%
\bibitem [{\citenamefont {Lee}\ \emph {et~al.}(2005)\citenamefont {Lee},
  \citenamefont {Brown}, \citenamefont {W. Yu}, \citenamefont {Naughton},\
  and\ \citenamefont {Chaikin}}]{Lee2005}%
  \BibitemOpen
  \bibfield  {author} {\bibinfo {author} {\bibfnamefont {I.}~\bibnamefont
  {Lee}}, \bibinfo {author} {\bibfnamefont {S.~E.}\ \bibnamefont {Brown}},
  \bibinfo {author} {\bibnamefont {W. Yu}}, \bibinfo {author} {\bibfnamefont
  {M.~J.}\ \bibnamefont {Naughton}}, \ and\ \bibinfo {author} {\bibfnamefont
  {P.~M.}\ \bibnamefont {Chaikin}},\ }\href@noop {} {\bibfield  {journal}
  {\bibinfo  {journal} {Phys. Rev. Lett.}\ }\textbf {\bibinfo {volume} {94}},\
  \bibinfo {pages} {197001} (\bibinfo {year} {2005})}\BibitemShut {NoStop}%
\bibitem [{\citenamefont {Podolsky}\ \emph {et~al.}(2004)\citenamefont
  {Podolsky}, \citenamefont {Altman}, \citenamefont {Rostunov},\ and\
  \citenamefont {Demler}}]{Podolsky2004}%
  \BibitemOpen
  \bibfield  {author} {\bibinfo {author} {\bibfnamefont {D.}~\bibnamefont
  {Podolsky}}, \bibinfo {author} {\bibfnamefont {E.}~\bibnamefont {Altman}},
  \bibinfo {author} {\bibfnamefont {T.}~\bibnamefont {Rostunov}}, \ and\
  \bibinfo {author} {\bibfnamefont {E.}~\bibnamefont {Demler}},\ }\href@noop {}
  {\bibfield  {journal} {\bibinfo  {journal} {Phys. Rev. Lett.}\ }\textbf
  {\bibinfo {volume} {93}},\ \bibinfo {pages} {246402} (\bibinfo {year}
  {2004})}\BibitemShut {NoStop}%
\bibitem [{\citenamefont {Zhang}\ and\ \citenamefont
  {de~Melo}(1999)}]{Zhang99}%
  \BibitemOpen
  \bibfield  {author} {\bibinfo {author} {\bibfnamefont {W.}~\bibnamefont
  {Zhang}}\ and\ \bibinfo {author} {\bibfnamefont {C.~S.}\ \bibnamefont
  {de~Melo}},\ }\href@noop {} {\bibfield  {journal} {\bibinfo  {journal} {J.
  Appl. Phys.}\ }\textbf {\bibinfo {volume} {97}},\ \bibinfo {pages} {10B108}
  (\bibinfo {year} {1999})}\BibitemShut {NoStop}%
\bibitem [{\citenamefont {Zegrodnik}\ and\ \citenamefont
  {Spalek}(2012)}]{Spalek}%
  \BibitemOpen
  \bibfield  {author} {\bibinfo {author} {\bibfnamefont {M.}~\bibnamefont
  {Zegrodnik}}\ and\ \bibinfo {author} {\bibfnamefont {J.}~\bibnamefont
  {Spalek}},\ }\href@noop {} {\bibfield  {journal} {\bibinfo  {journal} {Phys.
  Rev. B}\ }\textbf {\bibinfo {volume} {86}},\ \bibinfo {pages} {014505}
  (\bibinfo {year} {2012})}\BibitemShut {NoStop}%
\bibitem [{\citenamefont {Zhang}\ and\ \citenamefont
  {de~Melo}(2006)}]{Zhang2006}%
  \BibitemOpen
  \bibfield  {author} {\bibinfo {author} {\bibfnamefont {W.}~\bibnamefont
  {Zhang}}\ and\ \bibinfo {author} {\bibfnamefont {C.~A. R.~S.}\ \bibnamefont
  {de~Melo}},\ }\href@noop {} {\bibfield  {journal} {\bibinfo  {journal} {Phys.
  Rev. Lett.}\ }\textbf {\bibinfo {volume} {97}},\ \bibinfo {pages} {047001}
  (\bibinfo {year} {2006})}\BibitemShut {NoStop}%
\bibitem [{\citenamefont {Sigrist}\ and\ \citenamefont
  {Ueda}(1991)}]{sigrist91}%
  \BibitemOpen
  \bibfield  {author} {\bibinfo {author} {\bibfnamefont {M.}~\bibnamefont
  {Sigrist}}\ and\ \bibinfo {author} {\bibfnamefont {K.}~\bibnamefont {Ueda}},\
  }\href {\doibase 10.1103/RevModPhys.63.239} {\bibfield  {journal} {\bibinfo
  {journal} {Rev. Mod. Phys.}\ }\textbf {\bibinfo {volume} {63}},\ \bibinfo
  {pages} {239} (\bibinfo {year} {1991})}\BibitemShut {NoStop}%
\bibitem [{\citenamefont {Linder}\ and\ \citenamefont
  {Robinson}(2015)}]{Linder2015}%
  \BibitemOpen
  \bibfield  {author} {\bibinfo {author} {\bibfnamefont {J.}~\bibnamefont
  {Linder}}\ and\ \bibinfo {author} {\bibfnamefont {J.~W.~A.}\ \bibnamefont
  {Robinson}},\ }\href@noop {} {\bibfield  {journal} {\bibinfo  {journal} {Nat.
  Phys.}\ }\textbf {\bibinfo {volume} {11}},\ \bibinfo {pages} {307} (\bibinfo
  {year} {2015})}\BibitemShut {NoStop}%
\bibitem [{\citenamefont {Kwon}, \citenamefont {Sengupta},\ and\ \citenamefont
  {Yakovenko}(2004)}]{Yakovenko}%
  \BibitemOpen
  \bibfield  {author} {\bibinfo {author} {\bibfnamefont {H.}~\bibnamefont
  {Kwon}}, \bibinfo {author} {\bibfnamefont {K.}~\bibnamefont {Sengupta}}, \
  and\ \bibinfo {author} {\bibfnamefont {V.}~\bibnamefont {Yakovenko}},\
  }\href@noop {} {\bibfield  {journal} {\bibinfo  {journal} {Eur. Phys. J. B}\
  }\textbf {\bibinfo {volume} {37}},\ \bibinfo {pages} {349} (\bibinfo {year}
  {2004})}\BibitemShut {NoStop}%
\bibitem [{\citenamefont {Schnyder}\ \emph {et~al.}(2008)\citenamefont
  {Schnyder}, \citenamefont {Ryu}, \citenamefont {Furusaki},\ and\
  \citenamefont {Ludwig}}]{SRFL08}%
  \BibitemOpen
  \bibfield  {author} {\bibinfo {author} {\bibfnamefont {A.~P.}\ \bibnamefont
  {Schnyder}}, \bibinfo {author} {\bibfnamefont {S.}~\bibnamefont {Ryu}},
  \bibinfo {author} {\bibfnamefont {A.}~\bibnamefont {Furusaki}}, \ and\
  \bibinfo {author} {\bibfnamefont {A.~W.~W.}\ \bibnamefont {Ludwig}},\
  }\href@noop {} {\bibfield  {journal} {\bibinfo  {journal} {Phys. Rev. B}\
  }\textbf {\bibinfo {volume} {78}},\ \bibinfo {pages} {195125} (\bibinfo
  {year} {2008})}\BibitemShut {NoStop}%
\bibitem [{\citenamefont {Ryu}\ \emph {et~al.}(2010{\natexlab{a}})\citenamefont
  {Ryu}, \citenamefont {Schnyder}, \citenamefont {Furusaki},\ and\
  \citenamefont {Ludwig}}]{RSFL10}%
  \BibitemOpen
  \bibfield  {author} {\bibinfo {author} {\bibfnamefont {S.}~\bibnamefont
  {Ryu}}, \bibinfo {author} {\bibfnamefont {A.~P.}\ \bibnamefont {Schnyder}},
  \bibinfo {author} {\bibfnamefont {A.}~\bibnamefont {Furusaki}}, \ and\
  \bibinfo {author} {\bibfnamefont {A.}~\bibnamefont {Ludwig}},\ }\href@noop {}
  {\bibfield  {journal} {\bibinfo  {journal} {New J. Phys.}\ }\textbf {\bibinfo
  {volume} {12}},\ \bibinfo {pages} {065010} (\bibinfo {year}
  {2010}{\natexlab{a}})}\BibitemShut {NoStop}%
\bibitem [{\citenamefont {Qi}\ and\ \citenamefont {Zhang}(2011)}]{qi11}%
  \BibitemOpen
  \bibfield  {author} {\bibinfo {author} {\bibfnamefont {X.-L.}\ \bibnamefont
  {Qi}}\ and\ \bibinfo {author} {\bibfnamefont {S.-C.}\ \bibnamefont {Zhang}},\
  }\href@noop {} {\bibfield  {journal} {\bibinfo  {journal} {Rev. Mod. Phys.}\
  }\textbf {\bibinfo {volume} {83}},\ \bibinfo {pages} {1057} (\bibinfo {year}
  {2011})}\BibitemShut {NoStop}%
\bibitem [{\citenamefont {Tanaka}, \citenamefont {Sato},\ and\ \citenamefont
  {Nagaosa}(2012)}]{tanaka12}%
  \BibitemOpen
  \bibfield  {author} {\bibinfo {author} {\bibfnamefont {Y.}~\bibnamefont
  {Tanaka}}, \bibinfo {author} {\bibfnamefont {M.}~\bibnamefont {Sato}}, \ and\
  \bibinfo {author} {\bibfnamefont {N.}~\bibnamefont {Nagaosa}},\ }\href@noop
  {} {\bibfield  {journal} {\bibinfo  {journal} {J. Phys. Soc. Jpn.}\ }\textbf
  {\bibinfo {volume} {81}},\ \bibinfo {pages} {011013} (\bibinfo {year}
  {2012})}\BibitemShut {NoStop}%
\bibitem [{\citenamefont {Beenakker}(2013)}]{Beenakker13}%
  \BibitemOpen
  \bibfield  {author} {\bibinfo {author} {\bibfnamefont {C.~W.~J.}\
  \bibnamefont {Beenakker}},\ }\href@noop {} {\bibfield  {journal} {\bibinfo
  {journal} {Annu. Rev. Condens. Matter Phys.}\ }\textbf {\bibinfo {volume}
  {4}},\ \bibinfo {pages} {113} (\bibinfo {year} {2013})}\BibitemShut {NoStop}%
\bibitem [{\citenamefont {Leijnse}\ and\ \citenamefont
  {Flensberg}(2012)}]{Flensberg2012}%
  \BibitemOpen
  \bibfield  {author} {\bibinfo {author} {\bibfnamefont {M.}~\bibnamefont
  {Leijnse}}\ and\ \bibinfo {author} {\bibfnamefont {K.}~\bibnamefont
  {Flensberg}},\ }\href@noop {} {\bibfield  {journal} {\bibinfo  {journal}
  {Semiconductor Science and Technology}\ }\textbf {\bibinfo {volume} {27}},\
  \bibinfo {pages} {124003} (\bibinfo {year} {2012})}\BibitemShut {NoStop}%
\bibitem [{\citenamefont {Sato}\ and\ \citenamefont
  {Ando}(2017)}]{SatoAndo2017}%
  \BibitemOpen
  \bibfield  {author} {\bibinfo {author} {\bibfnamefont {M.}~\bibnamefont
  {Sato}}\ and\ \bibinfo {author} {\bibfnamefont {Y.}~\bibnamefont {Ando}},\
  }\href@noop {} {\bibfield  {journal} {\bibinfo  {journal} {Rep. Prog. Phys.}\
  }\textbf {\bibinfo {volume} {80}},\ \bibinfo {pages} {076501} (\bibinfo
  {year} {2017})}\BibitemShut {NoStop}%
\bibitem [{\citenamefont {Gentile}\ \emph {et~al.}(2013)\citenamefont
  {Gentile}, \citenamefont {Cuoco}, \citenamefont {Romano}, \citenamefont
  {Noce}, \citenamefont {Manske},\ and\ \citenamefont {Brydon}}]{Gentile13}%
  \BibitemOpen
  \bibfield  {author} {\bibinfo {author} {\bibfnamefont {P.}~\bibnamefont
  {Gentile}}, \bibinfo {author} {\bibfnamefont {M.}~\bibnamefont {Cuoco}},
  \bibinfo {author} {\bibfnamefont {A.}~\bibnamefont {Romano}}, \bibinfo
  {author} {\bibfnamefont {C.}~\bibnamefont {Noce}}, \bibinfo {author}
  {\bibfnamefont {D.}~\bibnamefont {Manske}}, \ and\ \bibinfo {author}
  {\bibfnamefont {P.~M.~R.}\ \bibnamefont {Brydon}},\ }\href@noop {} {\bibfield
   {journal} {\bibinfo  {journal} {Phys. Rev. Lett.}\ }\textbf {\bibinfo
  {volume} {111}},\ \bibinfo {pages} {097003} (\bibinfo {year}
  {2013})}\BibitemShut {NoStop}%
\bibitem [{\citenamefont {Terrade}, \citenamefont {Manske},\ and\ \citenamefont
  {Cuoco}(2016)}]{Terrade16}%
  \BibitemOpen
  \bibfield  {author} {\bibinfo {author} {\bibfnamefont {D.}~\bibnamefont
  {Terrade}}, \bibinfo {author} {\bibfnamefont {D.}~\bibnamefont {Manske}}, \
  and\ \bibinfo {author} {\bibfnamefont {M.}~\bibnamefont {Cuoco}},\
  }\href@noop {} {\bibfield  {journal} {\bibinfo  {journal} {Phys. Rev. B}\
  }\textbf {\bibinfo {volume} {93}},\ \bibinfo {pages} {104523} (\bibinfo
  {year} {2016})}\BibitemShut {NoStop}%
\bibitem [{\citenamefont {Mercaldo}, \citenamefont {Cuoco},\ and\ \citenamefont
  {Kotetes}(2016)}]{Mercaldo16}%
  \BibitemOpen
  \bibfield  {author} {\bibinfo {author} {\bibfnamefont {M.~T.}\ \bibnamefont
  {Mercaldo}}, \bibinfo {author} {\bibfnamefont {M.}~\bibnamefont {Cuoco}}, \
  and\ \bibinfo {author} {\bibfnamefont {P.}~\bibnamefont {Kotetes}},\
  }\href@noop {} {\bibfield  {journal} {\bibinfo  {journal} {Phys. Rev. B}\
  }\textbf {\bibinfo {volume} {94}},\ \bibinfo {pages} {140503(R)} (\bibinfo
  {year} {2016})}\BibitemShut {NoStop}%
\bibitem [{\citenamefont {Mercaldo}, \citenamefont {Cuoco},\ and\ \citenamefont
  {Kotetes}(2018)}]{Mercaldo18}%
  \BibitemOpen
  \bibfield  {author} {\bibinfo {author} {\bibfnamefont {M.~T.}\ \bibnamefont
  {Mercaldo}}, \bibinfo {author} {\bibfnamefont {M.}~\bibnamefont {Cuoco}}, \
  and\ \bibinfo {author} {\bibfnamefont {P.}~\bibnamefont {Kotetes}},\
  }\href@noop {} {\bibfield  {journal} {\bibinfo  {journal} {Physica B}\
  }\textbf {\bibinfo {volume} {536}},\ \bibinfo {pages} {730} (\bibinfo {year}
  {2018})}\BibitemShut {NoStop}%
\bibitem [{\citenamefont {Romano}\ \emph {et~al.}(2013)\citenamefont {Romano},
  \citenamefont {Gentile}, \citenamefont {Noce}, \citenamefont {Vekhter},\ and\
  \citenamefont {Cuoco}}]{Romano13}%
  \BibitemOpen
  \bibfield  {author} {\bibinfo {author} {\bibfnamefont {A.}~\bibnamefont
  {Romano}}, \bibinfo {author} {\bibfnamefont {P.}~\bibnamefont {Gentile}},
  \bibinfo {author} {\bibfnamefont {C.}~\bibnamefont {Noce}}, \bibinfo {author}
  {\bibfnamefont {I.}~\bibnamefont {Vekhter}}, \ and\ \bibinfo {author}
  {\bibfnamefont {M.}~\bibnamefont {Cuoco}},\ }\href@noop {} {\bibfield
  {journal} {\bibinfo  {journal} {Phys. Rev. Lett.}\ }\textbf {\bibinfo
  {volume} {110}},\ \bibinfo {pages} {267002} (\bibinfo {year}
  {2013})}\BibitemShut {NoStop}%
\bibitem [{\citenamefont {Romano}\ \emph {et~al.}(2016)\citenamefont {Romano},
  \citenamefont {Gentile}, \citenamefont {Noce}, \citenamefont {Vekhter},\ and\
  \citenamefont {Cuoco}}]{Romano16}%
  \BibitemOpen
  \bibfield  {author} {\bibinfo {author} {\bibfnamefont {A.}~\bibnamefont
  {Romano}}, \bibinfo {author} {\bibfnamefont {P.}~\bibnamefont {Gentile}},
  \bibinfo {author} {\bibfnamefont {C.}~\bibnamefont {Noce}}, \bibinfo {author}
  {\bibfnamefont {I.}~\bibnamefont {Vekhter}}, \ and\ \bibinfo {author}
  {\bibfnamefont {M.}~\bibnamefont {Cuoco}},\ }\href@noop {} {\bibfield
  {journal} {\bibinfo  {journal} {Phys. Rev. B}\ }\textbf {\bibinfo {volume}
  {93}},\ \bibinfo {pages} {014510} (\bibinfo {year} {2016})}\BibitemShut
  {NoStop}%
\bibitem [{\citenamefont {Romano}\ \emph {et~al.}(2017)\citenamefont {Romano},
  \citenamefont {Noce}, \citenamefont {Vekhter},\ and\ \citenamefont
  {Cuoco}}]{Romano17}%
  \BibitemOpen
  \bibfield  {author} {\bibinfo {author} {\bibfnamefont {A.}~\bibnamefont
  {Romano}}, \bibinfo {author} {\bibfnamefont {C.}~\bibnamefont {Noce}},
  \bibinfo {author} {\bibfnamefont {I.}~\bibnamefont {Vekhter}}, \ and\
  \bibinfo {author} {\bibfnamefont {M.}~\bibnamefont {Cuoco}},\ }\href@noop {}
  {\bibfield  {journal} {\bibinfo  {journal} {Phys. Rev. B}\ }\textbf {\bibinfo
  {volume} {96}},\ \bibinfo {pages} {054512} (\bibinfo {year}
  {2017})}\BibitemShut {NoStop}%
\bibitem [{\citenamefont {Manna}\ \emph {et~al.}(2017)\citenamefont {Manna},
  \citenamefont {Kamlapure}, \citenamefont {Cornils}, \citenamefont {Hanke},
  \citenamefont {Hedegaard}, \citenamefont {Bremholm}, \citenamefont {Iversen},
  \citenamefont {Hofmann}, \citenamefont {Wiebe},\ and\ \citenamefont
  {Wiesendanger}}]{Manna2017}%
  \BibitemOpen
  \bibfield  {author} {\bibinfo {author} {\bibfnamefont {S.}~\bibnamefont
  {Manna}}, \bibinfo {author} {\bibfnamefont {A.}~\bibnamefont {Kamlapure}},
  \bibinfo {author} {\bibfnamefont {L.}~\bibnamefont {Cornils}}, \bibinfo
  {author} {\bibfnamefont {T.}~\bibnamefont {Hanke}}, \bibinfo {author}
  {\bibfnamefont {E.~M.~J.}\ \bibnamefont {Hedegaard}}, \bibinfo {author}
  {\bibfnamefont {M.}~\bibnamefont {Bremholm}}, \bibinfo {author}
  {\bibfnamefont {B.~B.}\ \bibnamefont {Iversen}}, \bibinfo {author}
  {\bibfnamefont {P.}~\bibnamefont {Hofmann}}, \bibinfo {author} {\bibfnamefont
  {J.}~\bibnamefont {Wiebe}}, \ and\ \bibinfo {author} {\bibfnamefont
  {R.}~\bibnamefont {Wiesendanger}},\ }\href@noop {} {\bibfield  {journal}
  {\bibinfo  {journal} {Nat. Comm.}\ }\textbf {\bibinfo {volume} {8}},\
  \bibinfo {pages} {14074} (\bibinfo {year} {2017})}\BibitemShut {NoStop}%
\bibitem [{\citenamefont {Brzezicki}\ and\ \citenamefont
  {Cuoco}(2018)}]{Brzezicki2018}%
  \BibitemOpen
  \bibfield  {author} {\bibinfo {author} {\bibfnamefont {W.}~\bibnamefont
  {Brzezicki}}\ and\ \bibinfo {author} {\bibfnamefont {M.}~\bibnamefont
  {Cuoco}},\ }\href@noop {} {\bibfield  {journal} {\bibinfo  {journal} {Phys.
  Rev. B}\ }\textbf {\bibinfo {volume} {97}},\ \bibinfo {pages} {064513}
  (\bibinfo {year} {2018})}\BibitemShut {NoStop}%
\bibitem [{\citenamefont {Altland}\ and\ \citenamefont
  {Zirnbauer}(1997)}]{Altland}%
  \BibitemOpen
  \bibfield  {author} {\bibinfo {author} {\bibfnamefont {A.}~\bibnamefont
  {Altland}}\ and\ \bibinfo {author} {\bibfnamefont {M.~R.}\ \bibnamefont
  {Zirnbauer}},\ }\href@noop {} {\bibfield  {journal} {\bibinfo  {journal}
  {Phys. Rev. B}\ }\textbf {\bibinfo {volume} {55}},\ \bibinfo {pages} {1142}
  (\bibinfo {year} {1997})}\BibitemShut {NoStop}%
\bibitem [{\citenamefont {Kitaev}(2009)}]{KitaevClassi}%
  \BibitemOpen
  \bibfield  {author} {\bibinfo {author} {\bibfnamefont {A.}~\bibnamefont
  {Kitaev}},\ }\href@noop {} {\bibfield  {journal} {\bibinfo  {journal} {AIP
  Conf. Proc.}\ }\textbf {\bibinfo {volume} {1134}},\ \bibinfo {pages} {22}
  (\bibinfo {year} {2009})}\BibitemShut {NoStop}%
\bibitem [{\citenamefont {Ryu}\ \emph {et~al.}(2010{\natexlab{b}})\citenamefont
  {Ryu}, \citenamefont {Schnyder}, \citenamefont {Furusaki},\ and\
  \citenamefont {Ludwig}}]{Ryu}%
  \BibitemOpen
  \bibfield  {author} {\bibinfo {author} {\bibfnamefont {S.}~\bibnamefont
  {Ryu}}, \bibinfo {author} {\bibfnamefont {A.}~\bibnamefont {Schnyder}},
  \bibinfo {author} {\bibfnamefont {A.}~\bibnamefont {Furusaki}}, \ and\
  \bibinfo {author} {\bibfnamefont {A.}~\bibnamefont {Ludwig}},\ }\href@noop {}
  {\bibfield  {journal} {\bibinfo  {journal} {New. J. Phys.}\ }\textbf
  {\bibinfo {volume} {12}},\ \bibinfo {pages} {065010} (\bibinfo {year}
  {2010}{\natexlab{b}})}\BibitemShut {NoStop}%
\bibitem [{\citenamefont {Sato}\ \emph {et~al.}(2011)\citenamefont {Sato},
  \citenamefont {Tanaka}, \citenamefont {Yada},\ and\ \citenamefont
  {Yokoyama}}]{Sato2011PRB}%
  \BibitemOpen
  \bibfield  {author} {\bibinfo {author} {\bibfnamefont {M.}~\bibnamefont
  {Sato}}, \bibinfo {author} {\bibfnamefont {Y.}~\bibnamefont {Tanaka}},
  \bibinfo {author} {\bibfnamefont {K.}~\bibnamefont {Yada}}, \ and\ \bibinfo
  {author} {\bibfnamefont {T.}~\bibnamefont {Yokoyama}},\ }\href {\doibase
  10.1103/PhysRevB.83.224511} {\bibfield  {journal} {\bibinfo  {journal} {Phys.
  Rev. B}\ }\textbf {\bibinfo {volume} {83}},\ \bibinfo {pages} {224511}
  (\bibinfo {year} {2011})}\BibitemShut {NoStop}%
\end{thebibliography}
\end{document}